\documentclass[10 pt]{iopart}

\usepackage{epsfig}

\begin{document}

\title{Quantum Interferometry in $\rho^0$ Production in
Ultra-Peripheral Heavy Ion Collisions}

\author{Spencer R. Klein$^a$, for the STAR Collaboration}

\address{$^a$ Lawrence Berkeley National
Laboratory, Berkeley, CA 94720}

\begin{abstract}

In $\rho^0$ photoproduction in ultra-peripheral heavy ion collisions,
either ion can be the photon emitter or the target.  The two
possibilities are indistinguishable, and they should be able to
interfere, reducing $rho^0$ production at low transverse momentum,
$p_T<\hbar/\langle b\rangle$, where $\langle b\rangle$ is the median
impact parameter.   

The two $\rho^0$ production points are separated by $\langle b\rangle
\approx 18-46$ fm, while the $\rho^0$ decay before travelling 1 fm.
The two decay points are well separated in space-time, so the decays
proceed independently and any interference must involve the final
state $\pi^+\pi^-$.  This requires a non-local wave function.

\end{abstract}

\maketitle
\section{Introduction}

$\rho^0$ may be produced electromagnetically (via photoproduction) in
distant interactions between ultra-relativistic heavy ions.  In these
ultra-peripheral collisions (UPCs) \cite{reviews}, a photon from the
electromagnetic field of one nucleus fluctuates to a quark-antiquark
pair, which then scatters elastically from the other nucleus, emerging
as a vector meson \cite{phase}.  Elastic
scattering is colorless, and can be described in terms of soft Pomeron
exchange.  For light mesons like the $\rho^0$, the cross section rises
slowly with the photon energy $k$.  For small momentum transfers, the
elastic scattering is coherent over the entire nuclear target, greatly
increasing the cross section.  For heavy mesons, the scattering cross
section scales as the atomic number $A^2$, while for lighter mesons
like the $\rho^0$, a Glauber calculation finds that shadowing reduces
the dependence to approximately $A^{5/3}$ \cite{usPRC,strikmanblack}.

\epsfclipon
\begin{figure}
\begin{center}
{\setlength{\epsfxsize=1.9 in} 
\setlength{\epsfysize=1.9 in} 
\epsfbox{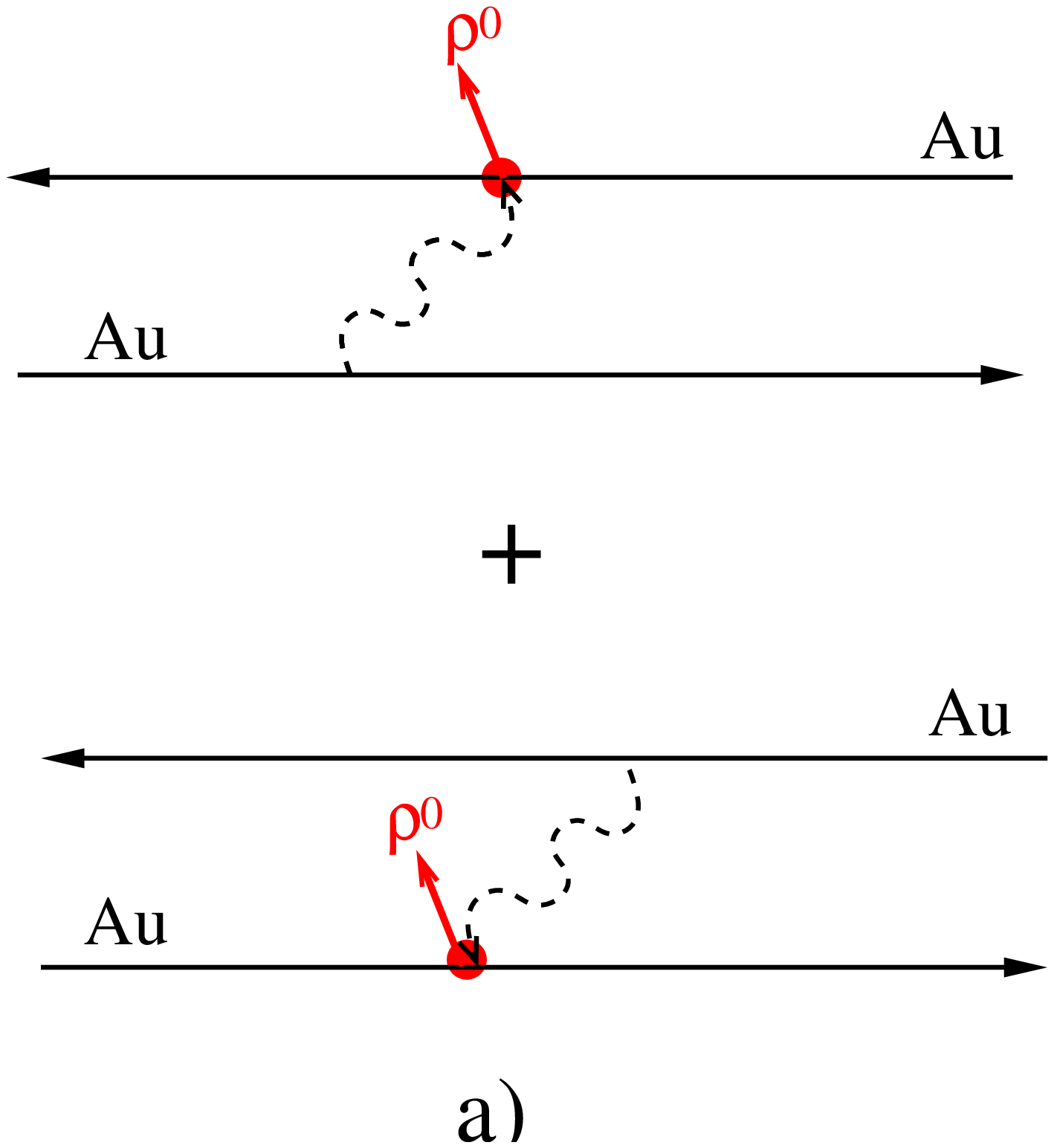}
\hskip .8 in
\setlength{\epsfxsize=1.9 in} 
\setlength{\epsfysize=1.9 in} 
\epsfbox{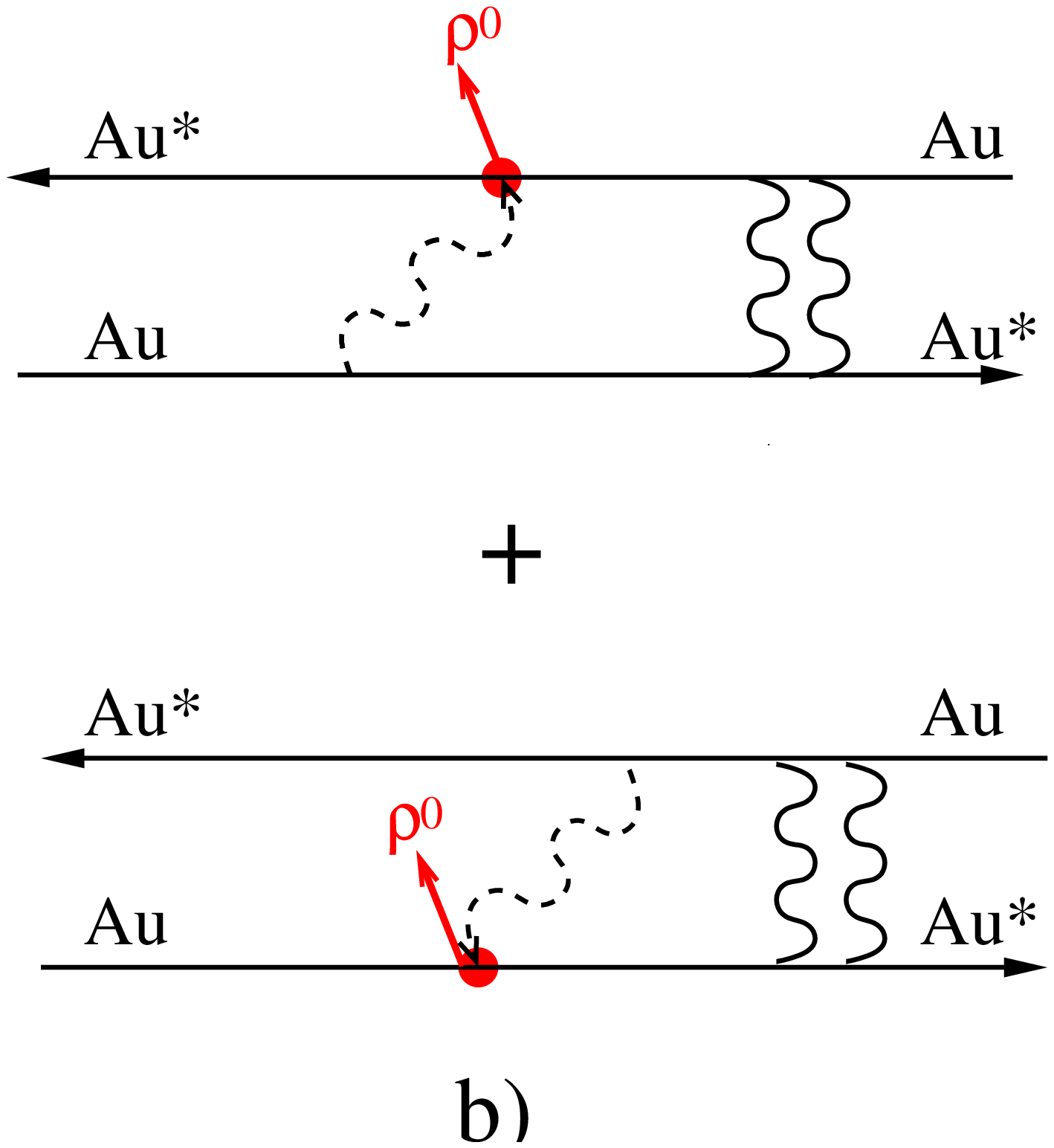}}
\end{center}
\caption[]{
\label{diagram}
(a) The two possibilities for exclusive $\rho^0$ production.
These two diagrams may be accompanied by mutual Coulomb excitation,
as in (b).  Different time orderings are not shown in (b), but do
not affect the interference.}
\end{figure}

The large photon flux and the coherent scattering lead to a large
cross section for $\rho^0$ production.  For 200 GeV per nucleon Au-Au
collisions at RHIC, the cross section for $\rho^0$ production is 590
mb, or about 8\% of the hadronic cross section \cite{usPRC}.  The STAR
collaboration has already measured $\rho^0$ production in Au-Au
collisions at an energy of 130 GeV per nucleon \cite{STARrho}.  The
cross sections agreed with the Glauber calculations.

The elastic scattering is mediated by the strong force, and so has a
short range \cite{muller}.  The $\rho$ production is localized to
within 1 fm of the two ions.  Figure 1a shows the two possibilities:
either nucleus 1 emits a photon which scatters off nucleus 2, or vice
versa.  These two possibilities are indistinguishable, and are related
by a parity transformation.  Vector mesons have negative parity, so
the two amplitudes subtract, with a transverse momentum ($p_T$)
dependent phase factor to account for the separation.  The cross
section is \cite{interfere}
\begin{equation}
\sigma = \big|A_1 - A_2\exp{(ip_T\cdot b)}\big|^2
\end{equation}
where $A_1$ and $A_2$ are the amplitudes for $\rho^0$ production from
the two directions.  At mid-rapidity $A_1=A_2$ and
\begin{equation}
\sigma = \sigma_0(b) \big[(1-\cos{(p_T\cdot b)}\big]
\label{eq:interfere}
\end{equation}
where $\sigma_0(b)$ is the cross section without interference.  The
system acts as a 2-slit interferometer, with slit separation $b$
\cite{prb}.  Of course, $b$ is unmeasurable, and the observed $p_T$
spectrum is obtained by integrating Eq. (1) over all $b$, subject to
the constraint that there are no hadronic interactions (roughly
equivalent to $b>2R_A$).  The cross section is suppressed for $p_T <
\hbar/\langle b\rangle$, where $\langle b\rangle$ is the median impact
parameter.  Interestingly, for photo

The $\rho$ decay distance, $\gamma\beta c\tau < 1$ fm is far less than
$\langle b\rangle\approx 46$ fm.  The $\rho^0$ decay before the
amplitudes from the two sources can overlap, making this an
interesting test of quantum mechanics \cite{physlett}.

The coupling constant $Z\alpha$ (where $\alpha\approx 1/137$ is the
electromagnetic coupling constant) is large, so multiple interactions
between a single ion pair are common.  the $\rho^0$ production may be
accompanied by mutual Coulomb excitation, as in Fig. 1b.  Each ion
emits a photon, which excites the other nucleus, usually to a giant
dipole resonance (GDR).  The resonance decays by emitting one or more
neutrons.  These neutrons are a clear experimental signature for
nuclear dissociation.

The mutual Coulomb dissociation also `tags' collisions with moderate
impact parameters, $2R_A < b < \approx 30$ fm \cite{baltzus}.  As long
as $k < \gamma\hbar/b$, the photon flux drops as $1/b^2$.  For a
single photon reaction like $\rho^0$ production, the production
probability $P(b)$ follows the same scaling.  For n-photon
interactions, $P(b)\approx 1/b^{2n}$ \cite{factorize} so the median
impact parameter is much smaller and interference is visible at larger
$p_T$.

Except for the shared impact parameter, the photon exchange reactions
are independent \cite{factorize}.  The cross section for $\rho^0$
production with mutual Coulomb dissociation is
\begin{equation}
\sigma = \int d^2b P_\rho(b) P_{2EXC}(b) [1-P_{had}(b)]
\label{eq:factorize}
\end{equation}
where $P_\rho(b)$ and $P_{2EXC}(b)$ are the probabilities for $\rho^0$
production and mutual excitation respectively. Here, $P_{had}(b)$ is
the probability of having a hadronic interaction; the last term has an
effect similar to setting a minimum impact parameter $b_{min} = 2
R_A$.  For gold ions, $P_\rho(2R_A)\approx 1\%$ and
$P_{2EXC}(2R_A)\approx 30\%$, so the probability of multiple
interactions is substantial.  Through the smaller $b$s, the photon
spectra in multi-photon interactions are harder than for single photon
reactions.

\section{The STAR Detector}

The Solenoidal Tracker at RHIC (STAR) studies heavy-ion and polarized
proton collisions at the RHIC collider Brookhaven National Laboratory.
STAR is optimized for high-multiplicity final states, but it is also
effective for 2-4 charged particle final states from UPCs.  STAR
collected UPC data in gold-gold collisions at energies of 130 GeV per
nucleon (in 2000) and 200 GeV per nucleon (in 2001).  STAR has also
studied $\rho^0$ photoproduction in 200 GeV per nucleon deuteron on
gold collisions \cite{falk}.  The $\rho^0$ interference analysis uses
the 2001 gold-gold collisions dataset.

Charged particles are reconstructed in a 4.2 meter long, 4 meter
diameter time projection chamber (TPC) \cite{TPC}.  This data was
taken with the TPC in a 0.5 T solenoidal magnetic field. The
$\pi^{\pm}$ reconstruction efficiency is high for tracks with
transverse momentum $p_T>100$ MeV/c and pseudorapidity $|\eta|<1.15$.
The track position and specific energy loss, $dE/dx$ were measured at
45 points for high momentum charged particles with $|\eta|<1$.  The
TPC is surrounded by 240 scintillator slats covering $|\eta|<1$,
comprising the central trigger barrel (CTB).  Two zero degree
calorimeters (ZDCs) are located 18 m upstream and downstream of the
interaction point.  These calorimeters are sensitive to neutrons from
mutual Coulomb dissociation \cite{ZDCs}.

Two different triggers \cite{trigger} were used to study interference.
The `topology' trigger selected events with 2 roughly back-to-back
tracks in the CTB.  It divided the CTB into 4 quadrants: north, south,
top and bottom, and required hits in the north and south quadrants.
The top and bottom regions were used as vetoes to reject cosmic rays.

The minimum bias trigger selected events with one or more neutrons in
each ZDC.  The ZDC signals were required to occur within 1 nsec,
restricting these events to the central 30 cm of the TPC.  Data from
both triggers was processed identically, except that events from the
CTB based trigger were distributed more broadly along the TPC axis,
and consequently, were accepted in a broader range.

\section{$\rho^0$ Production}

UPC $\rho^0$ production has a distinctive signature - two oppositely
charged tracks with small net $p_T$.  We select events with exactly 2
primary tracks that form a vertex.  The vertexing procedure considers
all of the tracks in an event, and rejects tracks that are
inconsistent with coming from a single vertex.  We allow a few
non-primary background tracks in the TPC. The topology and minimum
bias data were treated identically, except that allowance was made for
the different distribution of the accepted event production
points. This study used about 1.5 million minimum bias and about 1.7
million topology triggers.   This analysis will study the
$p_T$ spectra of produced $\rho^0$.  

Figure 2a shows the rapidity distribution of the coherent $\rho^0$
(pairs with $p_T < 150$ MeV/c).  The points (data) are in excellent
agreement with a calculation based on the soft Pomeron model and our
detector simulation.

Figure 2b shows the $\pi\pi$ invariant mass, $M_{\pi\pi}$ for the
sample.  The data is fit to a relativistic Breit-Wigner for the
$\rho^0$, plus a S\"oding term to account for the interference with
direct $\pi^+\pi^-$ production \cite{soding}.  The interference shifts
the peak of the distribution to lower $M_{\pi\pi}$. The direct
$\pi\pi$ to $\rho$ ratio agrees with the STAR 130 GeV analysis
\cite{STARrho} and with studies by the ZEUS collaboration in $\gamma
p$ interactions \cite{ZEUS}.  However, with the nuclear coherence, the
STAR data is at smaller $|t|$ than the ZEUS analysis, so the two
results may not be directly comparable \cite{marks}.

\begin{figure}
\setlength{\epsfxsize=2.7 in} 
\setlength{\epsfysize=2.0 in} 
\epsfbox{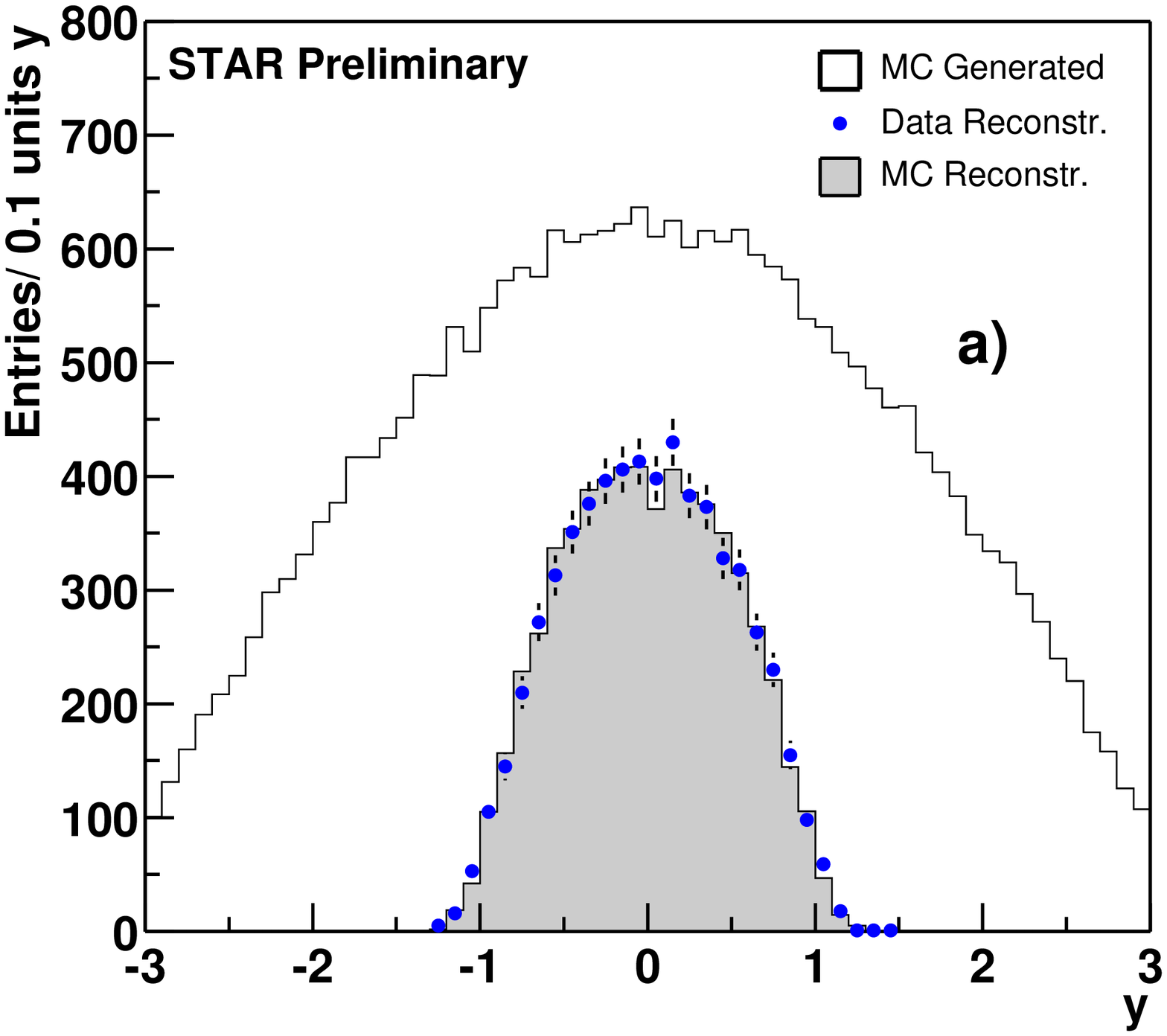}
\setlength{\epsfxsize=2.7 in} 
\setlength{\epsfysize=2.0 in} 
\epsfbox{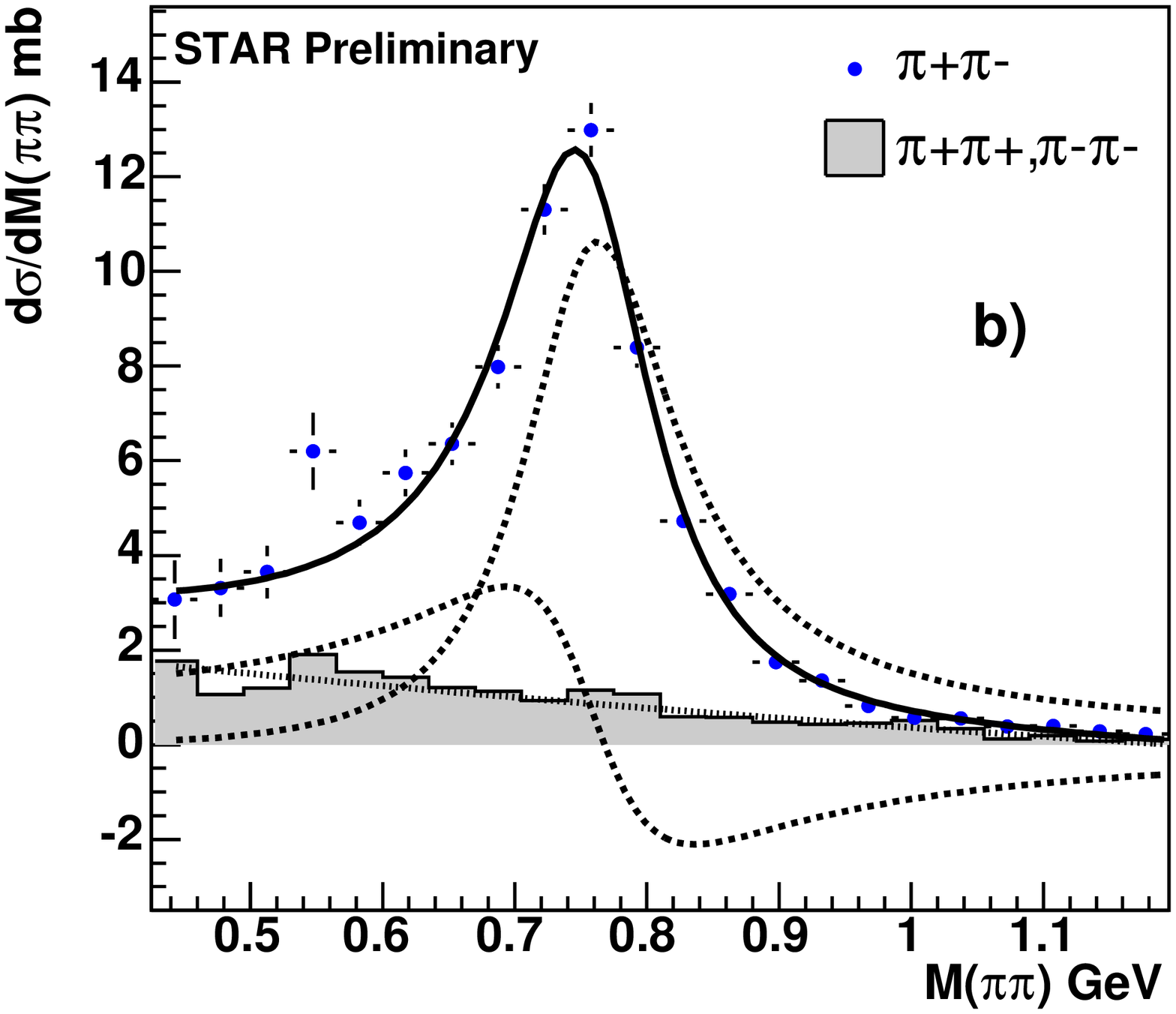}
\caption[]{
\label{rhoplots}
The (a) rapidity and (b) $\pi\pi$ invariant mass spectra of $\rho^0$
using data from the minimum bias trigger.  The mass is fit to to a
Breit-Wigner function for the $\rho^0$, plus a {S\"oding} term due to
interference with a direct $\pi\pi$ component.  The combinatoric
hadronic background is also shown.}
\end{figure}

\section{$\rho^0$ Interferometry}

Stringent event selection criteria were used to select a clean,
low-background sample for the $\rho^0$ interference study.  Events
were required to have exactly two tracks with a vertex within 50 (100)
cm longitudinally of the center of the TPC for the minimum bias
(topology trigger) sample.  $\rho^0$ candidates with $|y|<0.1$ were
eliminated to avoid possible contamination from cosmic rays.  The
pairs were required to have 550 MeV $< M_{\pi\pi} < 920$ MeV.  The
minimum mass cut removed most of the hadronic backgrounds, which are
concentrated at low $M_{\pi\pi}$.  These cuts reduced the like-sign
pair background to a few percent. With these cuts, the background from
misidentified two-photon reactions like $\gamma\gamma\rightarrow
e^+e^-$ should be very small.  The sample still includes direct pions,
which should have the same spin/parity and quantum mechanical behavior
as the pion pairs from $\rho$ decay.  We do not distinguish between
the two sources.

The magnitude of the interference depends on the ratio of the
amplitudes for $\rho$ production on the two nuclei.  Away from $y=0$,
the photon energies for the two photon directions differ, $k_{1,2} =
M_V/2 \exp(\pm y/2)$; the amplitudes differ and the interference is
sub-maximal.  Although it is not expected in the soft-Pomeron model,
the the photon energy difference could introduce a small $\rho^0$
production phase difference, which could affect the interference
\cite{phase}.  This analysis focuses on the region near mid-rapidity
where any phase difference should be small.  A Monte Carlo is used to
find the interference for different rapidity ranges
\cite{usPRC,interfere}.

We study the $p_T$ spectra using the variable $t_\perp = p_T^2$. At
RHIC energies, the longitudinal component of the 4-momentum transfer
is small, so $t\approx t_\perp$.  Without interference, the spectrum
$dN/dt \propto\exp{(-bt)}$ \cite{interfere,ting}.

Figure 3 compares the uncorrected minimum bias data for $0.1 <
|\eta|<0.5$ with two simulations, with and without interference.  Both
simulations include the detector response.  The data has a significant
downturn for $t<0.001$ GeV$^2$, consistent with the $\langle b\rangle
= 18$ fm expected for a $\rho^0$ accompanied by mutual
excitation \cite{factorize}. This data matches calculation with
interference, but not the no-interference result.

\epsfclipon
\begin{figure}
\label{rawinterfere}
\setlength{\epsfxsize=3.0 in} 
\setlength{\epsfysize=2.1 in} 
\centerline{\epsffile{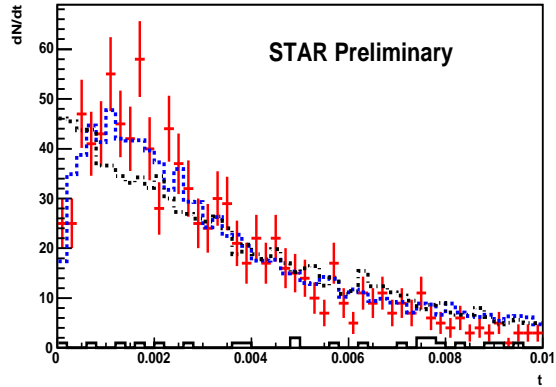}}
\caption[]{Raw (uncorrected) $t_\perp$ spectrum for $\rho^0$ sample
for $0.1 < |y|<0.5$ for the topology data.  The points (red) are the
data.  The dashed histogram (blue) is a simulation that assumes that
there is interference, while the dot-dash histogram (black) is based
on a calculation without interference.  The solid black histogram with
very few counts is the like-sign background.}
\end{figure}

The efficiency corrected data are shown in Fig. 4. Minimum bias and
topology data are shown separately, each with two rapidity bins: $0.1
< |y| < 0.5$ and $0.5 < |y| < 1.0$.  The efficiency is independent of
$p_T$.  However $p_T$ smearing (resolution) does affect the spectrum
slightly; the efficiency correction removes this.  The $\rho^0$ $p_T$
resolution is about 9 MeV/c, while the 1st $t$ bin covers 0 to (15
MeV/c)$^2$.  Interference depletes the first few bins, but feed down
from the higher t bins partially repopulates them.

\begin{figure*}
\label{tspectra}
\setlength{\epsfxsize= 2.9 in} 
\setlength{\epsfysize=2.1 in}
\centerline{\epsffile{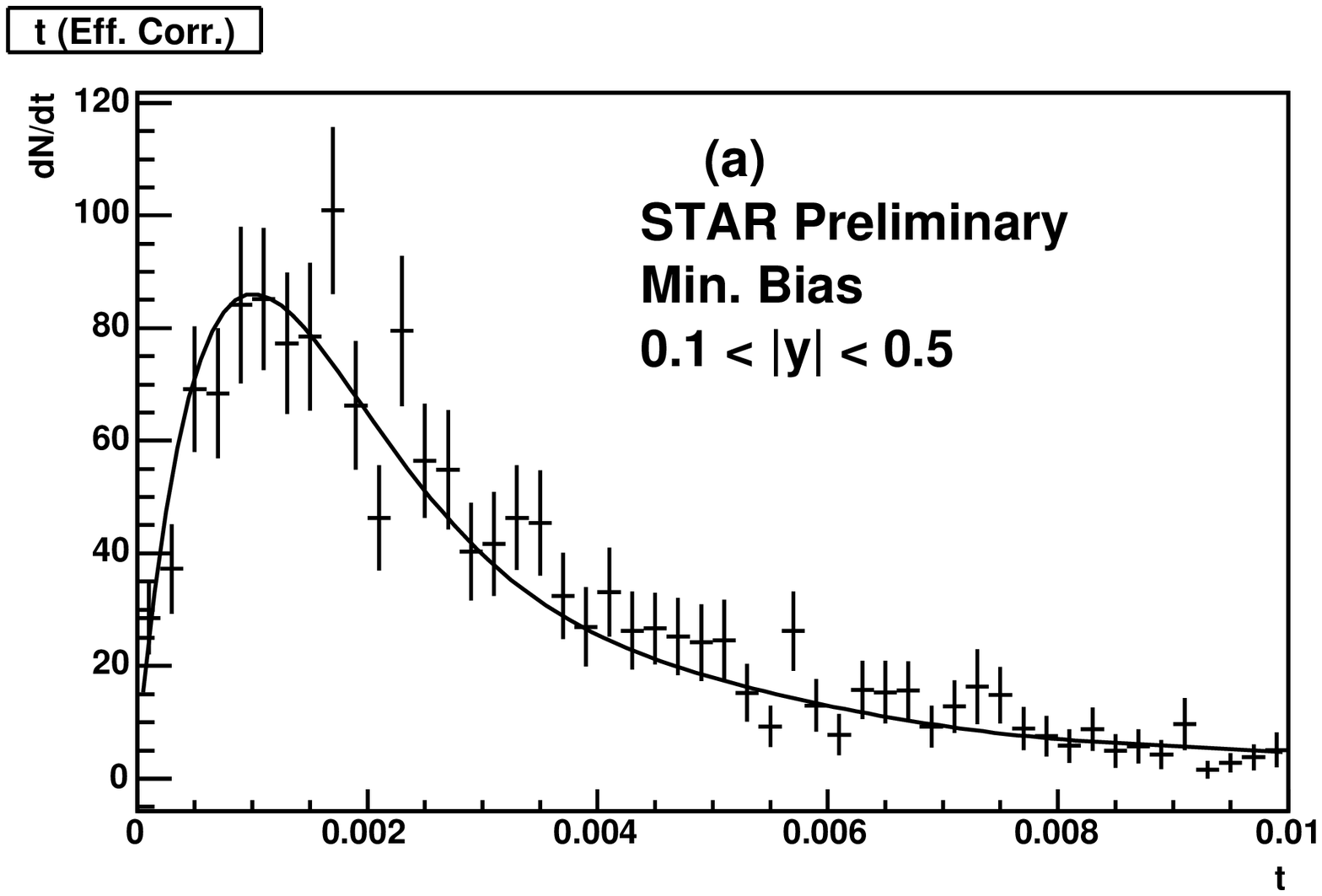}
\setlength{\epsfxsize= 2.9 in} 
\setlength{\epsfysize=2.1 in}
\epsffile{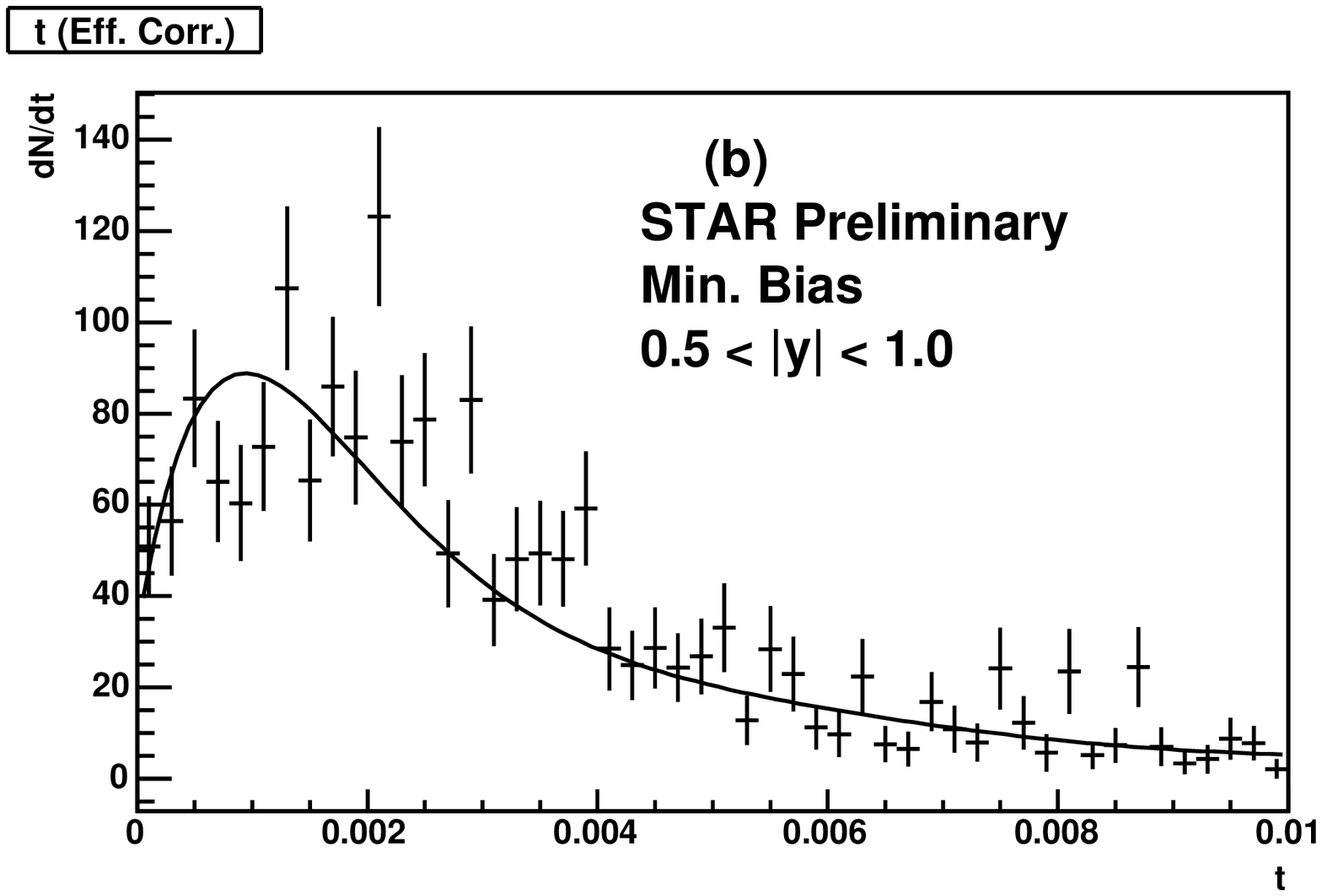}}
\vskip .01 in
\setlength{\epsfxsize= 2.9 in} 
\setlength{\epsfysize=2.1 in}
\centerline{\epsffile{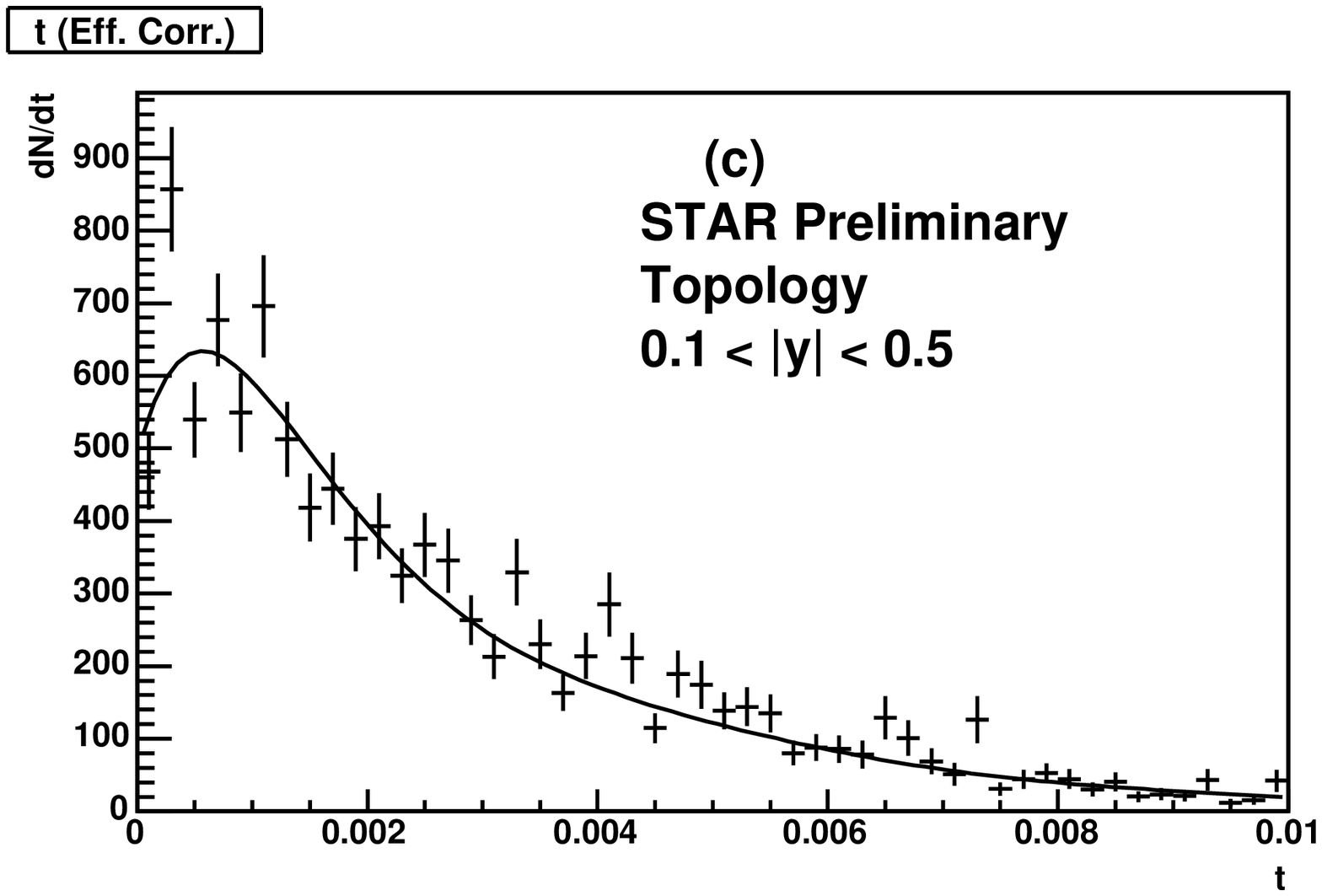}
\setlength{\epsfxsize= 2.9 in} 
\setlength{\epsfysize=2.1 in}
\epsffile{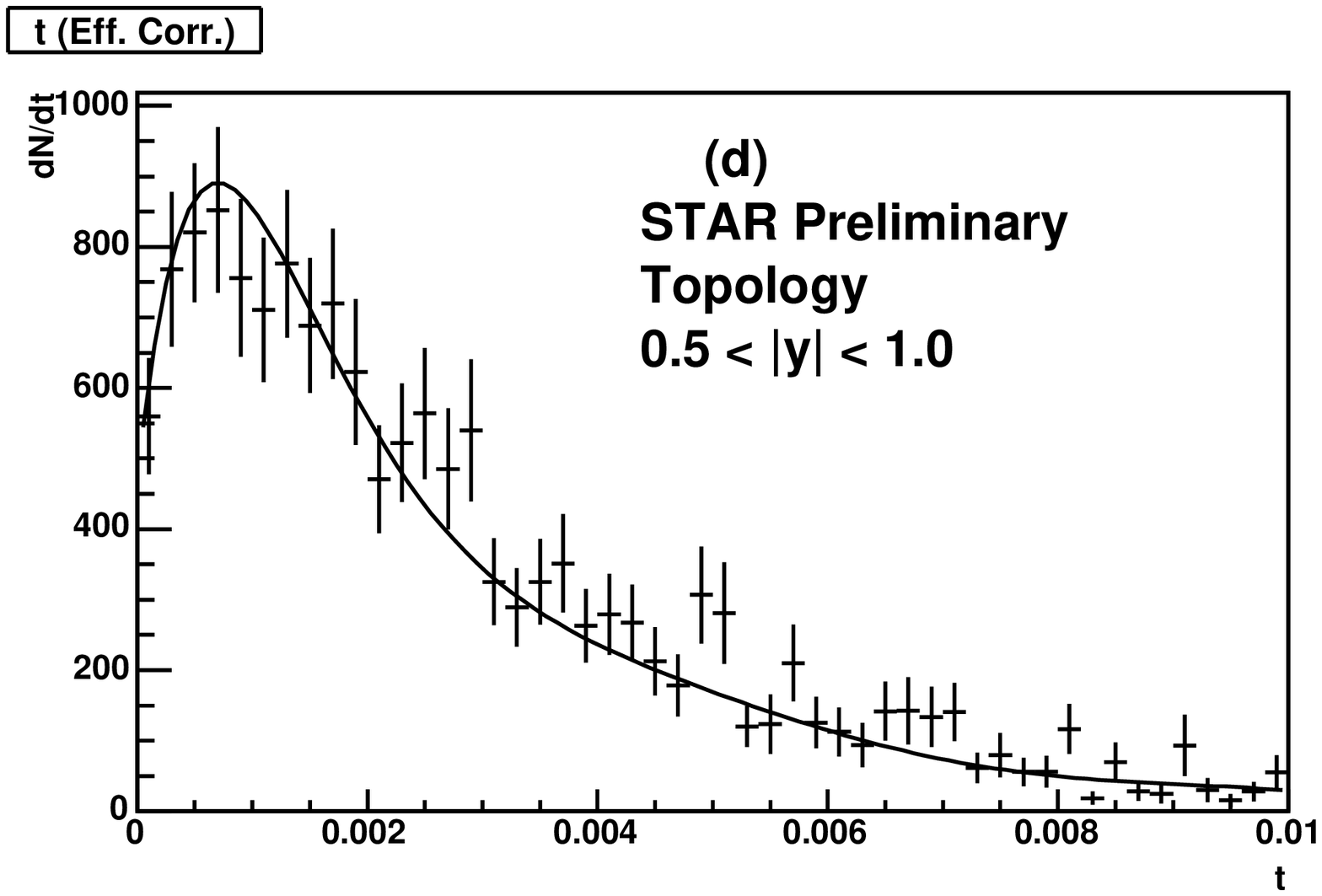}}
\caption[]{Efficiency corrected $t_\perp$ spectrum for $\rho^0$ from
(a) mutual dissociation with $0.1 < |y|<0.5$, (b) mutual dissociation
with $0.5 < |y|<1.0$, (c) topology trigger with $0.1 < |y|<0.5$ and
(d) topology trigger with $0.5 < |y|<1.0$.  The points are the data,
while the solid line is the result of the fit to Eq. 4.}
\end{figure*}

The data is fit to the 3-parameter form:
\begin{equation}
{dN\over dt} = a \exp(-bt) [1+ c(R(t)-1]
\end{equation}
where $R(t) = Int(t)/Noint(t)$ is the ratio of the Monte Carlo
t-spectra with and without interference.  Here, $a$ is the overall
normalization, the slope $b\approx R_A^2$, and $c$ is the degree of
spectral modification; $c=0$ corresponds to no interference while
$c=1$ is the calculated interference.  This functional form separates
the interference ($c$) from the nuclear form factor ($b$).

Table 1 gives the results of the fits.  At small-rapidities the
amplitudes from the two directions are similar and the interference
reduces the cross section at $p_T=0$ by more than at larger
rapidities.  In the minimum bias data, the interference extends to
higher $p_T$ than the topology data because the former has a smaller
$\langle b\rangle$.

\begin{table}
\begin{center}
\begin{tabular}{lcrrr}
Trigger & 
$|$Rapidity$|$ & 
\multicolumn{1}{c} b (GeV$^{-2}$) & 
\multicolumn{1}{c} c & 
\multicolumn{1}{c} \ $\chi^2/$DOF \\
\hline
Minimum Bias & $0.1 - 0.5$&  $301\pm14$  & $1.01\pm 0.08$ & 50/47\\
Minimum Bias & $0.5 - 1.0$&  $304\pm 15$ & $0.78\pm 0.13$ &73/47 \\
Topology & $0.1 - 0.5$&  $361\pm 10$ & $0.71\pm0.16$  & 81/47    \\
Topology & $0.5 - 1.0$&  $368\pm 12$ & $1.22\pm0.21$ & 50/47     \\
\hline
\end{tabular}
\end{center}
\caption[]{Results of the fits to the two datasets for the two
rapidity bins.}
\label{t:fitout}

\end{table}

The 4 $c$ values are consistent within errors; the weighted
average is $c = 0.93 \pm 0.06$.  The $b$ values for the
minimum bias and exclusive $\rho^0$ data differ by 20\%: $364\pm7$
GeV$^{-2}$ for the exclusive $\rho$ versus $303\pm 10$ GeV$^{-2}$
for the Coulomb breakup events.  

The different $b$ values may be attributed to the different impact
parameter distributions caused by the nuclear breakup tagging in the
minimum bias data.  The photon flux depends on the impact parameter as
$1/b^2$.  When $b\approx {\rm few} R_A$, $\rho$ are more likely to be
produced on the side of the target near the photon emitter than on the
far side.  $\rho^0$ production is concentrated on the near side,
leading to a smaller effective production volume and the smaller $b$.
This near-side skewing affects the interference slightly, but is not
included in current calculations.

Systematic errors come from a variety of sources.  We have studied the
effect of detector distortions (primarily $p_T$ smearing) by turning
off the detector simulation and comparing raw simulations with
reconstructed data.  This lowered $c$ by 18\%; if the detector
simulation is 75\% correct (a very conservative assumption), then
detector effects are a less than 5\% systematic uncertainty. Minor
systematic uncertainties come from backgrounds and the fitting
procedure.  We have compared the data and simulations for a variety of
kinematic and detector-based variables, and found good agreement. We
estimate a preliminary experimental systematic uncertainty of 8\%.

This analysis depends on the shape of $R(t)$, which is calculated
following Ref. \cite{interfere}.  There are some simplifications in
Ref. \cite{interfere}.  The calculation averages the photon flux over
the surface of the nucleus, rather than incorporating the proper
$1/b^2$ weighting.  This may partly explain the relatively poor fit
for the large-$|y|$ Coulomb breakup fit.  We estimate that the
effective impact parameter should be within 10\% of the actual ion-ion
separation for the coulomb breakup data, and 3\% for the exclusive
$\rho$.  The uncertainties in the calculations should be at most a
15\% effect. With this, the interference is $93\pm6 (stat.) \pm 8
(syst.) \pm 15 (theory) \%$ (STAR preliminary) of that expected.

The $\rho$ decays rapidly, with $\gamma\beta c\tau\ll \langle
b\rangle$ and the two $\rho$ decay points are well separated in
space-time.  In the usual space-time picture, the decays occur
independently, and any interference must involve the final state
$\pi^+\pi^-$.  Interference must involve identical final states from
the two sources.  However, given the large available phase space for
the decays, this is very unlikely for independent decays.

One interpretation of this result is given in Ref. \cite{physlett}. In
it, the interference occurs because the post-decay wave function
includes amplitudes for all possible final states, then the amplitudes
for identical states subtract, and the interference is visible.
Because of the two sources, the $\pi^+\pi^-$ wave function is
non-factorizable, and thus exhibits the Einstein-Podolsky-Rosen
paradox \cite{EPR}.  We find that the decoherence, $1-c$, due to
environmental or other factors is less than 43\% at the 90\%
confidence level.

\section{Conclusions}

The STAR collaboration has studied exclusive $\rho^0$ photoproduction
and $\rho^0$ photoproduction accompanied by nuclear excitation.  The
cross sections and rapidity distributions match the predictions of the
soft Pomeron model.

Production at low$-p_T$ is suppressed, as expected if $\rho^0$
production at the two ions interferes destructively.  The 
suppression is $93 \pm 6 (stat.) \pm 8 (syst.) \pm 15 (theory)
\%$ of that expected.  The decoherence is less than $43\%$
at a 90\% confidence level.

We thank Tony Baltz and Joakim Nystrand for help with the theoretical
calculations.  This work was supported by the U.S. DOE under contract
number DE-AC-03-76SF00098.


\section{References}

\begin{thebibliography}{99}
\def\etal{{\it et al.}}

\bibitem{reviews}G. Baur {\it et al.}, Phys. Rep. {\bf 364}, 359
(2002); F. Krauss, M. Greiner and G. Soff, Prog. Part. Nucl.
Phys. {\bf 39}, 503 (1997).

\bibitem{phase}T. H. Baur {\it et al.}, Rev. Mod. Phys.  {\bf 50}, 261
(1978).

\bibitem{usPRC}S. Klein and J. Nystrand, Phys. Rev. {\bf C60}, 014903
(1999).

\bibitem{strikmanblack} L. Frankfurt, M. Strikman and M. Zhalov,
Phys. Rev. {\bf C67}, 034901 (2003).

\bibitem{STARrho}C. Adler {\it et al.}, Phys. Rev. Lett. {\bf 89},
272302 (2002).

\bibitem{muller}B. M\"uller and A. J. Schramm, Nuclear
Physics {\bf A523}, 677 (1991).

\bibitem{interfere}S. Klein and J. Nystrand, Phys. Rev. Lett.
{\bf 84}, 2330 (2000).

\bibitem{prb} An analagous two-slit interferometer is described by
T. Sudbery in {\it Quantum Concepts in Space and Time}, ed. R. Penrose
and C. J. Isham, (Oxford, 1986).

\bibitem{physlett}S. Klein and J. Nystrand, Phys. Lett. {\bf A308},
323 (2003).  

\bibitem{baltzus}A. Baltz, S. Klein and J. Nystrand,
Phys. Rev. Lett. {\bf 89}, 012301 (2002).

\bibitem{factorize}G. Baur {\it et al.}, Nucl. Phys. {\bf A729}, 787
(2003).

\bibitem{falk}F. Meissner and V. B. Morozov, nucl-ex/0307006 (2003).

\bibitem{TPC}M. Anderson \etal, Nucl. Instrum. \& Meth. {\bf A499}, 659
(2003); M. Anderson \etal, Nucl. Instrum. \& Meth. {\bf A499}, 679
(2003).

\bibitem{ZDCs}C. Adler {\it et al.}, Nucl. Instrum. \& Meth.  {\bf
A470}, 488 (2001).

\bibitem{trigger}F. S. Bieser {\it et al.}, Nucl. Instrum. \&
Meth. {\bf A499}, 766 (2003).

\bibitem{soding}P. S\"oding, Phys. Lett. {\bf 19}, 702 (1966).

\bibitem{ZEUS}J. Breitweg {\it et al.}, Eur. Phys. J.
{\bf C15}, 1 (2000).

\bibitem{marks}Mark Strikman, private communication (2003).

\bibitem{ting}M. Alvensleben {\it et al.}, Phys. Rev. Lett.
{\bf 24}, 792 (1970).

\bibitem{EPR}A. Einstein, B. Podolsky and N. Rosen, Phys. Rev.
{\bf 47}, 777 (1935).

\end{thebibliography}
\end{document}